\documentclass[letterpaper, 10 pt, conference]{IEEEtran}

\usepackage{amsfonts,amsmath,amsthm,amscd,amssymb,array}

\usepackage{MnSymbol}

\usepackage[T1]{fontenc}
\usepackage[latin1]{inputenc}

\usepackage{nicefrac}				
\usepackage{fancyhdr,xcolor,fancybox}
\usepackage{algpseudocode,algorithm}		
\usepackage{makeidx} 				
\usepackage{pstricks,pst-plot,pst-node} 	
\usepackage{graphicx}				
\usepackage{listings}				
\usepackage{subfigure}				
\usepackage{enumerate}				
\usepackage[intoc,english]{nomencl}		
\usepackage{acronym}				
\usepackage{cite}				
\usepackage{ifsym}
\usepackage{xspace}
\usepackage{ulem}
\usepackage{array}
\usepackage{tabularx}
\usepackage{flushend}				
\usepackage{marvosym}
\usepackage{tipa} 				

\newcommand{\reffigure}[1]{Fig.~\ref{#1}}

\newtheorem{theorem}{Theorem}					
\newtheorem{lemma}[theorem]{Lemma}				

\newcommand{\ie}{\mbox{i.\,e.}\xspace}				
\newcommand{\Wlog}{\mbox{w.\,l.\,o.\,g.}\xspace}		
\newcommand{\resp}{\mbox{respectively}\xspace}			
\newcommand{\etal}{\mbox{et\,al.}\xspace}			

\newcommand{\Tx}{\mathrm{T}}					
\newcommand{\Rx}{\mathrm{R}}					
\renewcommand{\mod}{\mathrm{mod}}				
\newcommand{\transpose}{\mathsf{T}}				
\newcommand{\DoF}{\mathrm{DoF}}					
\newcommand{\modxn}{\ \mod\,(x^n-1)}				

\IEEEoverridecommandlockouts

\begin{document}
\addtolength{\voffset}{0.3cm}
\title{$Y$-\,$\Delta$ Product in $3$-Way $\Delta$ and $Y$-\,Channels \\ for Cyclic Interference and Signal Alignment}

\author{
\IEEEauthorblockN{
Henning Maier\IEEEauthorrefmark{1},
Anas Chaaban\IEEEauthorrefmark{2} and
Rudolf Mathar\IEEEauthorrefmark{1}}

\IEEEauthorblockA{\IEEEauthorrefmark{1}Institute for Theoretical Information Technology, RWTH Aachen University, 52056 Aachen, Germany}
\IEEEauthorblockA{\IEEEauthorrefmark{2}Chair of Communication Systems, Ruhr-Universit\"at Bochum, 44780 Bochum, Germany}

Email: \{maier, mathar\}@ti.rwth-aachen.de, \{anas.chaaban\}@rub.de

%
\thanks{This work has been supported by the \emph{Deutsche Forschungs\-gemein\-schaft} (DFG) within the project \emph{Power
Adjustment and Constructive Interference Alignment for Wireless Networks} (PACIA - Ma 1184/15-2) of the DFG program
\emph{Communication in Interference Limited Networks} (COIN) and furthermore by the UMIC Research Centre, RWTH Aachen University.}
}

\maketitle
\IEEEpeerreviewmaketitle

\begin{abstract}
In a full-duplex $3$-way $\Delta$ channel, three transceivers communicate to each other, so that a number of six messages is exchanged.
In a \mbox{$Y$-\,channel}, however, these three transceivers are connected to an intermediate full-duplex relay.
Loop-back self-interference is suppressed perfectly.
The relay forwards network-coded messages to their dedicated users by means of interference alignment (IA) and signal alignment.
A conceptual channel model with cyclic shifts described by a polynomial ring is considered for these two related channels.
The maximally achievable rates in terms of the degrees of freedom measure are derived.
We observe that the $Y$-\,channel and the  $3$-way $\Delta$ channel provide a $Y$-$\Delta$ product relationship.
Moreover, we briefly discuss how this analysis relates to spatial IA and MIMO IA.
\end{abstract}

\vspace{-2mm}
\section{Introduction}
In two-way full-duplex communication systems, users operate both as transmitters and receivers, \ie, transceivers, and exchange messages with each other in a bidirectional manner.
A generalization of the two-way channel is the $K$-user full-duplex interference channel.
In \cite{318}, this channel is considered for time-varying channel coefficients.
It is shown that a full-duplex channel can be equivalently represented by a fully-connected $K$-\,user interference channel with perfect feedback links between the transmitters and the receivers with the same index.
To achieve the upper bounds on the degrees of freedom (DoF), the innovative concept of Interference Alignment (IA) \cite{001} is applied.
For $K=3$ we call this a $3$-way $\Delta$ channel.

A \mbox{$Y$-\,channel}~\cite{539} is a related $3$-way communication system but with one intermediate relay.
Each transceiver sends two messages to the relay, and the relay forwards three network-coded messages back to the dedicated users.
The DoF of the MIMO $Y$-\,channel with an arbitrary antenna configuration are provided in~\cite{582}.
In~\cite{437}, the capacity region of the related linear shift deterministic \mbox{$Y$-\,channel} is derived.

A conceptual channel model based on polynomials and inspired by cyclic codes as introduced in \cite{Z3a} to investigate the impact of interference in multi-user networks.
Therein, \emph{Cyclic IA} on the \mbox{$X$-\,channel} and the $K$-\,user interference channel is considered.
A Cyclic Interference Neutralization scheme on this channel model was investigated in~\cite{Z4a}.
The polynomial model is closely related to the finite-field model in \cite{597} and to the linear shift deterministic channel model introduced in~\cite{005}.

\textbf{Contributions.}
In \mbox{$3$-way $\Delta$} channels and \mbox{$Y$-\,channels}, each user intends to convey two messages, i.e., one message to each other.
We derive optimal Cyclic~IA schemes for both channels in terms of the conceptual polynomial channel model.
We observe that the provided schemes achieving the same proposed upper bounds and are essentially equivalent.
The \mbox{$Y$-channel} is expressed by a $\Delta$ channel using a \mbox{$Y$-\,$\Delta$} product relationship.
This relationship is evidently motivated by elementary circuit theory.
To the best of our knowledge, it has not been reported in terms of information theory of multi-user communications yet.
Note that, in contrast to our previous works \cite{Z3a} and \cite{Z4a}, the channel matrices are not subject to further conditions.

\textbf{Organization.}
In Sec.\,\ref{sec:sysmod} we define the conceptual model of the polynomial representation for the \mbox{$3$-way $\Delta$} channel and the $Y$-\,channel.
An upper bound on the DoF is provided in Sec.\,\ref{sec:Obounds}.
We propose corresponding Cyclic~IA schemes for both channels in Sec.\,\ref{sec:3uTW} and~\ref{sec:Y}.
The $Y$-\,$\Delta$ product of Cyclic~IA is discussed in Sec.\,\ref{sec:equivalence}.
In Sec.\,\ref{sec:mimo-dof}, we briefly relate our results to IA schemes in~\cite{582,318}.

\vspace{-0.5mm}
\section{System Model}
\label{sec:sysmod}
We adapt the notation used in \cite{Z3a} and \cite{Z4a}.
The set of user-indices is defined by~\mbox{$\mathcal{K}:=\{1,2,3\}$}.
In a full-duplex system, a user is a combined transmitter and receiver and denoted as a transceiver~$\Tx_i$, $i \in \mathcal{K}$.
There are $6$ independent message vectors $\boldsymbol{w}_{ji}$, namely, $\boldsymbol{w}_{12}$, $\boldsymbol{w}_{21}$, $\boldsymbol{w}_{13}$, $\boldsymbol{w}_{31}$, $\boldsymbol{w}_{23}$ and $\boldsymbol{w}_{32}$, dedicated to be conveyed from a  transceiver~$\Tx_i$ to a transceiver~$\Tx_j$, with $i \neq j \in \mathcal{K}$, \ie, each transceiver broadcasts two message vectors.
The message vectors $\boldsymbol{w}_{ji}$ contain $\alpha_{ji} \in \mathbb{N}$ submessages $W_{ji}^{[*]}$ and are denoted by a vector \mbox{$\boldsymbol{w}_{ji} = (W_{ji}^{[1]}, \dots, W_{ji}^{[\alpha_{ji}]})$}.
A submessage is a string of $t \in \mathbb{N}$ bits $W_{ji}^{[*]} \in \mathbb{B}=\{0,1\}^t$.
We interpret the different number of the submessages as individual rate demands per user-pair.
The number of submessage per dedicated user-pair is expressed by the messaging matrix:
\begin{align}
 \label{eqn:msg-matrix}
 \boldsymbol{M} &=
  \left(
  \begin{array}{ccc}
    0 & \alpha_{12} & \alpha_{13} \\
    \alpha_{21} & 0 & \alpha_{23} \\
    \alpha_{31} & \alpha_{32} & 0 \\
  \end{array}
  \right),
\end{align}
and the total number of submessages amounts to:
\begin{align*}
 M = \left\Vert\boldsymbol{M}\right\Vert_1 = \alpha_{12} + \alpha_{21} + \alpha_{13} + \alpha_{31} + \alpha_{23} + \alpha_{32}.
\end{align*}
We consider polynomial rings $\mathbb{F}(x)/(x^n-1)$ with the indeterminate $x$.
The channel access at each $\Tx_i$ is partitioned into \mbox{$n \in \mathbb{N}$} dimensions and each single dimen\-sion has length one.
A single dimension within the $n$ dimensions is addressed by an offset $x^0, x^1,\dots, x^{n-1}$, from $0$ (no offset) to $n-1$ (maximal offset).
A transmitted signal $u(x) \in \mathbb{F}(x)/(x^n-1)$ is a polynomial where submessages are allocated to a subset of the coefficients.
E.g., a cyclic shift of a polynomial $u(x)=Wx^l$ by $k$ positions is expressed by~$x^k u(x) \equiv Wx^{k+l}\,\modxn$.
For notational brevity, we will mostly omit the modulo~$x^n-1$ in congruences. 

\subsection{$3$-Way $\Delta$ Channel}

\begin{figure}[t]
 \centering
 \includegraphics[width=78mm]{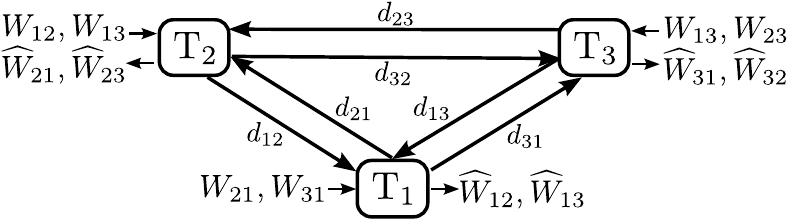}
 \caption{The (upside-down) $3$-way $\Delta$ channel with three transceivers $\mathrm{T}_1, \mathrm{T}_2$ and $\mathrm{T}_3$ has six independent messages $W_{ji}$ transmitted and six corresponding estimated messages $\widehat{W}_{ji}$ received by the dedicated transceivers.
 The influence of the channel is parameterized by a corresponding~$d_{ji}$.
 }
 \label{fig:channel}
 \vspace{-0.6cm}
\end{figure}
The setup of the $3$-way $\Delta$ channel is depicted in \reffigure{fig:channel}.
The signal transmitted from $\Tx_i$ is represented by a polynomial $u_{i}(x)$, with messages allocated to distinct offset parameters  \mbox{$p_{ji}^{[m]} \in \mathbb{N}$}:
\begin{align}
 u_i(x) &\equiv \sum\nolimits_{j=1,j \neq i}^3 u_{ji}(x), \\
 u_{ji}(x) &\equiv \sum\nolimits_{m=1}^{\alpha_{ji}} W_{ji}^{[m]} x^{p_{ji}^{[m]}}.
\end{align}
The channel matrix is defined by \mbox{$\boldsymbol{D}=(d_{ji})_{1\leq j,i \leq 3}$} with independent elements \mbox{$d_{ji} \in \mathcal{D}:=\{x^k \mid k \in \mathbb{N}\}$}.
The received signal at $\Tx_j$, $j \in \mathcal{K}$, is a superposition of shifted polynomials~$u_i(x)$:
\begin{align}
 r_j(x) \equiv \sum\nolimits_{i=1}^{3}d_{ji}u_{i}(x). 
\end{align}
In a vector notation, we can compactly express this as:
\begin{align}
 \boldsymbol{r}^\transpose \equiv \boldsymbol{D}\boldsymbol{u}^\transpose.
\end{align}
with $\boldsymbol{r}=(r_1(x),r_2(x),r_3(x))$, $\boldsymbol{u}=(u_1(x),u_2(x),u_3(x))$.
The congruence symbol indicates that each element is reduced modulo~$x^n-1$.

\subsection{$Y$-\,Channel}

\begin{figure}[t]
 \centering
 \includegraphics[width=78mm]{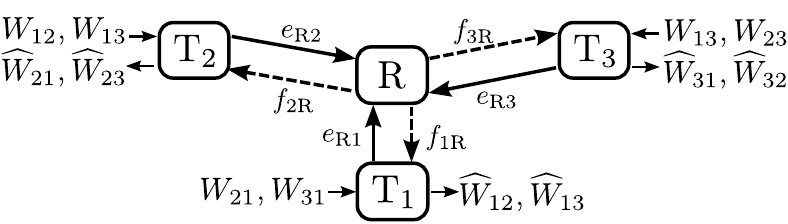}
 \caption{The $Y$-\,channel with three transceivers $\mathrm{T}_1, \mathrm{T}_2$ and $\mathrm{T}_3$ and one intermediate relay $\mathrm{R}_1$, has six independent messages $W_{ji}$ transmitted and six estimated messages $\widehat{W}_{ji}$ received by the dedicated transceivers. The influence of the channel is parameterized by the corresponding~$e_{\Rx i}$ and~$f_{j \Rx}$. The solid lines describe the uplink and the dashed lines the downlink.}
 \label{fig:channel-Y}
 \vspace{-0.6cm}
\end{figure}

The setup of the \mbox{$Y$-\,channel} is depicted in Fig.~\ref{fig:channel-Y}.
In this case, we include an intermediate full-duplex relay~$\Rx$.
All transceivers are linked to $\Rx$ and there is no direct link between the three transceivers $\Tx_1,\Tx_2,\Tx_3$.

The uplink phase denotes the transmission in the first hop from the transceivers $\Tx_i$ to the relay $\Rx$.
The uplink is a channel vector \mbox{$\boldsymbol{e}=(e_{\mathrm{R}1},e_{\mathrm{R}2},e_{\mathrm{R}3})$}, with $e_{\mathrm{R}1},e_{\mathrm{R}2},e_{\mathrm{R}3} \in \mathcal{D}$.
The received signal at relay $\Rx$~is:
\begin{align}
 r_{\Rx}(x) &\equiv \boldsymbol{eu}^\transpose \equiv \sum\nolimits_{i=1}^{3} e_{\Rx i} u_i(x).
\end{align}
The downlink phase denotes the transmission in the second hop from the single relay $\Rx$ to the three transceivers $\Tx_j$ and the downlink channel vector is denoted by \mbox{$\boldsymbol{f}=(f_{1\mathrm{R}},f_{2\mathrm{R}},f_{3\mathrm{R}})$} with $f_{1\mathrm{R}},f_{2\mathrm{R}},f_{3\mathrm{R}} \in \mathcal{D}$.
The relay forwards its received signals to all transceivers:
\begin{align}
 \boldsymbol{r}^\transpose &\equiv \boldsymbol{f}^\transpose r_{\Rx}(x) \equiv (f_{1R},f_{2R},f_{3R})^\transpose \,r_{\Rx}(x).
\end{align}
The two channel vectors $\boldsymbol{e}$ and $\boldsymbol{f}$ are independent w.l.o.g.
Due to the full-duplex assumption for each transceiver and relay, both phases are performed simultaneously in each time-step.
But to maintain causality, the second hop is delayed by one time-step to the corresponding first hop.
In case of instantaneous relays, these two hops would not be delayed.

\section{Upper Bound on the Degrees of Freedom}
\label{sec:Obounds}
We use the Degrees of Freedom (DoF) to measure the achieved rate.
It is defined by the number  of interference-free submessages $M$ conveyed within $n$ dimensions~\cite{Z3a}:
\begin{align}
  \label{eqn:DoF}
 \mathrm{DoF} = \frac{M}{n}.
\end{align}
In analogy to \cite[Thm. 1]{059b}, and applied in terms of the CPCM, the DoF for general $K_{\mathrm{T}} \times K_{\mathrm{R}}$ $X$\,-\,channels with $K_{\mathrm{T}}$ transmitters and $K_{\mathrm{R}}$ receivers are upper bounded by \cite{Z3a}:
\begin{align}
 \label{eqn:obound-gen}
 \DoF &\leq \frac{\sum\nolimits_{j=1}^{K_{\mathrm{R}}}\sum\nolimits_{i=1}^{K_{\mathrm{T}}} \alpha_{ji}}
 {\underset{\alpha_{ji}}{\max}\left(\sum\nolimits_{i=1}^{K_{\mathrm{T}}} \alpha_{ji} + \sum\nolimits_{j=1}^{K_{\mathrm{R}}} \alpha_{ji}-\alpha_{ji}\right)},
\end{align}
for a $K_{\mathrm{T}} \times K_{\mathrm{R}}$ messaging matrix $\boldsymbol{M}$ with non-zero elements on the diagonal.
However, in a $K$-\,user multi-way channel we consider $K_{\mathrm{T}} = K_{\mathrm{R}} = K$ and a zero-diagonal in~$\boldsymbol{M}$, since the transmitters $\mathrm{Tx}_i$ and receivers $\mathrm{Rx}_i$ are pair-wise co-located into one combined transceiver~$\Tx_i$.
As the corresponding upper bound does not include the link from a transceiver to itself, the $\mathrm{max}$-operation is not taken over the diagonal elements:
\begin{align}
 \label{eqn:obound-TW}
 \DoF &\leq \frac{\sum\nolimits_{j=1}^{K}\sum\nolimits_{i=1, i \neq j}^{K} \alpha_{ji}}
 {\underset{\alpha_{ji}, i \neq j}{\max}\left(\sum\nolimits_{i=1, i \neq j}^{K} \alpha_{ji} + \sum\nolimits_{j=1, i \neq j}^{K} \alpha_{ji}-\alpha_{ji}\right)}.
\end{align}
The denominator is also a lower bound on the dimensions~$n$.

\section{Cyclic IA on the $3$-Way $\Delta$ Channel}
\label{sec:3uTW}
\subsection{Separability Conditions}
We define a set of separability conditions \cite{Z3a} necessary for an interference-free communication in a multi-user two-way channel with pair-wise distinct \mbox{$i,j,k \in \mathcal{K}$}.
In multi-way channels, back-propagated self-interference is known a priori and removed by Self-Interference Cancellation (SIC), so that signals over loop-back links $d_{ii}, i \in \mathcal{K}$ are discarded.

As multiple desired signals must remain decodable at each \mbox{dedicated} receiver, the \emph{multiple-access interference conditions} must be satisfied with \mbox{$m \in \{1,\dots,\alpha_{ji}]\}$}, \mbox{$m^\prime \in \{1,\dots,\alpha_{jl}\}$}:
  \begin{align}
    \label{eqn:mac-FD}
    d_{ji} x^{p_{ji}^{[m]}} \nequiv d_{jl}x^{p_{jl}^{[m^\prime]}}.
  \end{align}
Note that the indices can be relabelled to obtain these two conditions equivalent to \eqref{eqn:mac-FD} for corresponding $m$, $m^\prime$:
\begin{align}
  d_{ki}x^{p_{ki}^{[m]}} &\nequiv d_{kj}x^{p_{kj}^{[m^\prime]}}, \label{eqn:mac-FD2} \\
  d_{ij}x^{p_{ij}^{[m]}} &\nequiv d_{ik}x^{p_{ik}^{[m^\prime]}}. \label{eqn:mac-FD3}
\end{align}

Furthermore, multiple signals transmitted from the same user, but dedicated for different receivers, must be separable at the transmitter-side, \ie, they must satisfy the following \emph{intra-user interference conditions} with \mbox{$m \in \{1,\dots,\alpha_{ji}\}$}, \mbox{$m^\prime \in \{1,\dots,\alpha_{ki}\}$}:
  \begin{align}
   \label{eqn:intra-FD3}
   x^{p_{ji}^{[m]}} \nequiv x^{p_{ki}^{[m^\prime]}}.
  \end{align}
And by relabelling the indices, we equivalently obtain:
\begin{align}
 x^{p_{ij}^{[m]}} &\nequiv x^{p_{kj}^{[m^\prime]}}. \label{eqn:intra-FD3b}
\end{align}
Interfering signals that are not dedicated for a given receiver must satisfy the following \emph{inter-user interference conditions} with \mbox{$m \in \{1,\dots,\alpha_{ji}\}$}, \mbox{$m^\prime \in \{1,\dots,\alpha_{ki}\}$}:
  \begin{align}
  \label{eqn:inter-FD}
   d_{ji} x^{p_{ji}^{[m]}} \nequiv d_{jl}x^{p_{kl}^{[m^\prime]}}.
  \end{align}

\subsection{Elementary Case with $M=6$ messages}
Firstly, we consider an elementary case with messaging matrix \mbox{$\boldsymbol{M}=\boldsymbol{1}_{3 \times 3} - \boldsymbol{I}_3$}.
Each $\Tx_i$ has only $\alpha_{ji}=1$ submessage per message, so that we may omit the superscript notation for now.
We apply the basic idea of IA, \ie, to combine and overlap all interfering signals within the smallest possible set of dimensions at each receiver.
Accordingly, we propose \emph{Cyclic~IA} for the corresponding two interference signals:
\begin{align}
 d_{ji}x^{p_{ki}} \equiv d_{jk}x^{p_{ik}}\,\mod(x^3-1), \label{eqn:cyclicIA-FD}
\end{align}
with pair-wise distinct $i,j,k \in \mathcal{K}$.
Such a scheme is called perfect Cyclic~IA since the interference overlaps exactly in the same number of dimensions.
Relabelling the indices provides:
\begin{align}
  d_{kj}x^{p_{ij}} \equiv d_{ki}x^{p_{ji}}\,\mod(x^3-1), \label{eqn:cyclicIA-FD2} \\
  d_{ik}x^{p_{jk}} \equiv d_{ij}x^{p_{kj}}\,\mod(x^3-1). \label{eqn:cyclicIA-FD3}
\end{align}

\begin{lemma}
 \label{lem:FD}
 Perfect Cyclic~IA achieves the upper bound of $2$~DoF on the $3$-way $\Delta$ channel.
\end{lemma}
\emph{Proof:}
 \begin{itemize}
  \item [(a)] \emph{Necessity of $n \geq 3$ dimensions}: \newline
  A number of $n=3$ dimensions is necessary, since the two dedicated signals must occupy one dimension each by \eqref{eqn:mac-FD} to be decodable and the interference signals are aligned to one dimension at each receiver to satisfy \eqref{eqn:intra-FD3}.
  \item [(b)] \emph{Sufficiency of Cyclic~IA with $n=3$}:  \newline
  We fix the parameter~$p_{ik}$.
  Then, \eqref{eqn:cyclicIA-FD} yields a unique~$p_{ki}$.
  By \eqref{eqn:mac-FD2}, only two valid solutions remain for~$p_{kj}$,
  and by~\eqref{eqn:mac-FD3}, only two valid solutions remain for~$p_{ij}$.
  W.l.o.g. we may choose one valid solution for $p_{ij}$, satisfying a relabelled version of \eqref{eqn:mac-FD3}.
  Now, a unique solution for~$p_{ji}$ is provided by~\eqref{eqn:cyclicIA-FD2}.
  Since $p_{ij}$ and $p_{ki}$ are already fixed, only one unique solution for $p_{kj}$ remains by \eqref{eqn:intra-FD3} and \eqref{eqn:mac-FD2}.
  The parameter $p_{jk}$ yields from \eqref{eqn:cyclicIA-FD3}.
  All parameters $p_{ik},p_{ki},p_{ij},p_{ji},p_{jk},p_{kj}$ are fixed.

  It is still to check whether the inter-user interference conditions in \eqref{eqn:inter-FD} hold.
  By using \eqref{eqn:cyclicIA-FD}, the condition \eqref{eqn:inter-FD} can be simplified to \eqref{eqn:intra-FD3b}.
  As \eqref{eqn:intra-FD3b} is equivalent to \eqref{eqn:intra-FD3} which already holds, all separability conditions are satisfied.
  Here, Cyclic IA is independent of~$\boldsymbol{D}$ and always achieves the upper~bound of $\frac{M}{n}=\frac{6}{3}=2$ DoF.~$\blacksquare$
\end{itemize}

\subsection{General Case}
In the case of the general messaging matrix given by~\eqref{eqn:msg-matrix}, the inter-user and intra-user interference is aligned in pairs for the indices \mbox{$m \in \{1,\dots,\min(\alpha_{ki},\alpha_{ik})\}$}~by:
\begin{align}
 \label{eqn:alignFDgen}
 d_{ji}x^{p_{ki}^{[m]}} \equiv d_{jk}x^{p_{ik}^{[m]}}\,\modxn
\end{align}
the remaining submessages are transmitted by multiple-access.
\begin{theorem}
 \label{thm:IA-FD-arb}
 Cyclic~IA on the $3$-way $\Delta$ channel achieves the upper bound:
 \begin{align*}
  \DoF \leq \frac{\alpha_{12}+\alpha_{21}+\alpha_{13}+\alpha_{31}+\alpha_{23}+\alpha_{32}}{\max(n_1,n_2,n_3)} \leq 2,
 \end{align*}
 within $n \geq \max(n_1,n_2,n_3)$ dimensions for distinct \mbox{$i,j,k \in \mathcal{K}$}:
 \begin{align*}
   n_j &= \alpha_{ji}+\alpha_{jk}+\max(\alpha_{ik},\alpha_{ki}).
 \end{align*}
\end{theorem}
\emph{Proof:}
\begin{itemize}
 \item [(a)] \emph{Necessity of $n \geq \max(n_1,n_2,n_3)$ dimensions:} \newline
 The denominator of the upper bound \eqref{eqn:obound-TW} yields:
 \begin{align*}
  n \geq \max(&\alpha_{32}+\alpha_{13}+\alpha_{12}, \alpha_{12}+\alpha_{13}+\alpha_{23}, \nonumber \\
	      &\alpha_{21}+\alpha_{23}+\alpha_{31}, \alpha_{21}+\alpha_{23}+\alpha_{13}, \nonumber \\
	      &\alpha_{21}+\alpha_{31}+\alpha_{32}, \alpha_{31}+\alpha_{32}+\alpha_{12}).
 \end{align*}
 By rewriting this in terms of distinct indices $i,j,k \in \mathcal{K}$, we obtain the constraints on $n$ provided in the theorem.
 \item [(b)] \emph{Sufficiency of Cyclic~IA:} \newline
 The scheme of Lem.\,\ref{lem:FD} is repeated for each parameter-pair with the same index \mbox{$m=1,\dots,\min(\alpha_{ik},\alpha_{ki})$} within \mbox{$\min(\alpha_{ik},\alpha_{ki})$} dimensions.
 Each aligned pair $m$ occupies exactly one dimen\-sion at each transceiver and thus the separability conditions hold.

 At $\Tx_j$, $\min(\alpha_{ik},\alpha_{ki})$ dimensions already contain the aligned interference.
 The interference by the \mbox{$|\alpha_{ik}-\alpha_{ki}|$} remaining submessages from either $\Tx_i$ or $\Tx_k$, are separately allocated in a multiple-access scheme to the yet unused dimensions at $\Tx_j$ and demand $|\alpha_{ik}-\alpha_{ki}|$ dimensions to satisfy the separability conditions.
 Now, the interference spaces span \mbox{$\min(\alpha_{ik},\alpha_{ki})+|\alpha_{ik}-\alpha_{ki}|$} \mbox{$=\max(\alpha_{ik},\alpha_{ki})$} dimensions.
 Altogether, each dedicated message is conveyed interference-free within a number of \mbox{$n_j = \alpha_{ji}+\alpha_{jk}+\max(\alpha_{ik},\alpha_{ki})$} dimensions at $\Tx_j$.

  The DoF are maximized if interference is perfectly aligned.
  Hence, if $\alpha_{23}=\alpha_{32}:=\alpha_{1}, \alpha_{13}=\alpha_{31}:=\alpha_{2}$ and $\alpha_{12}=\alpha_{21}:=\alpha_{3}$ hold, we obtain $n=n_1=n_2=n_3=\alpha_1+\alpha_2+\alpha_3$ so that at most $\frac{2(\alpha_1+\alpha_2+\alpha_3)}{\alpha_1+\alpha_2+\alpha_3}=2$ DoF are achievable.~$\blacksquare$
\end{itemize}

\section{Cyclic~SA on the $Y$-\,Channel}
\label{sec:Y}

\subsection{Separability Conditions}
The separability conditions of the $Y$-\,channel are closely related to \eqref{eqn:mac-FD}, \eqref{eqn:intra-FD3} and \eqref{eqn:inter-FD}.
For distinct indices $i,j,k \in \mathcal{K}$ and the indices $m,m^\prime$ defined in relation to the given parameters, the particular separability conditions of the \mbox{$Y$-\,channel} are:
\begin{itemize}
 \item \emph{multiple-access interference conditions}:
\begin{align}
 \label{eqn:mac-if-1R}
 e_{\mathrm{R}i} f_{j \Rx}x^{p_{ji}^{[m]}} &\nequiv e_{\mathrm{R}k} f_{j\Rx} x^{p_{jk}^{[m^\prime]}}, \nonumber \\
 \Rightarrow e_{\mathrm{R}i} x^{p_{ji}^{[m]}} &\nequiv e_{\mathrm{R}k} x^{p_{jk}^{[m^\prime]}},
\end{align}
 \item \emph{intra-user interference conditions}:
\begin{align}
 \label{eqn:intra-if-1R}
 x^{p_{ji}^{[m]}} \nequiv x^{p_{ki}^{[m^\prime]}},
\end{align}
 \item \emph{inter-user interference conditions}:
\begin{align}
 \label{eqn:inter-if-1R}
 e_{\mathrm{R}i} f_{j \Rx} x^{p_{ji}^{[m]}} &\nequiv e_{\mathrm{R}k} f_{j \Rx} x^{p_{ik}^{[m^\prime]}},  \nonumber \\
 \Rightarrow
 e_{\mathrm{R}i} x^{p_{ji}^{[m]}} &\nequiv e_{\mathrm{R}k} x^{p_{ik}^{[m^\prime]}}.
\end{align}
\end{itemize}

\subsection{Elementary Case with $M=6$ messages}
Again, we consider the messaging matrix \mbox{$\boldsymbol{M}=\boldsymbol{1}_{3 \times 3} - \boldsymbol{I}_3$}.
In the uplink, the dedicated signal from $\Tx_i$ to $\Tx_j$, \mbox{$i \neq j \in \mathcal{K}$}, is aligned at $\Rx$ to the dedicated signal from~$\Tx_j$ to $\Tx_i$, i.e., to the signal dedicated for the index-swapped direction:
\begin{align}
\label{eqn:IA-1R}
 e_{\mathrm{R}i} x^{p_{ji}} \equiv e_{\mathrm{R}j} x^{p_{ij}}\,\mod(x^3-1).
\end{align}
In the downlink, the messages back-propagated from relay $\mathrm{R}$ to the original transceiver are known a priori and removed by SIC.
Channel vector $\boldsymbol{f}$ may be chosen arbitrarily, since the separability conditions in \eqref{eqn:mac-if-1R} to \eqref{eqn:inter-if-1R} do not impose any constraints on $\boldsymbol{f}$ at all.
\begin{lemma}
 \label{lem:Y-perfIA}
 A perfect Cyclic~SA scheme with SIC achieves $2$~DoF on the $Y$-channel.
\end{lemma}
\emph{Proof:}
\begin{itemize}
 \item [(a)] \emph{Necessity of $n \geq 3$ dimensions:} \newline
  As in Lem.\,\ref{lem:FD}, one dimensions for each dedicated signal and one dimension for inter and intra-user interference is necessary to satisfy the separability conditions.
 \item [(b)] \emph{Sufficiency of Cyclic IA and SIC:} \newline
  In the uplink phase, we fix a parameter $p_{ji}$ w.l.o.g.
  Then, $p_{ij}$ is determined uniquely by~\eqref{eqn:IA-1R}.
  By further satisfying \eqref{eqn:intra-if-1R}, $p_{ki}$ can be chosen from two possible solutions.
  Then, $p_{ik}$ is also determined uniquely by a relabelled version of~\eqref{eqn:IA-1R}: \mbox{$e_{\mathrm{R}i} x^{p_{ki}} \equiv e_{\mathrm{R}k} x^{p_{ik}}$}.
  Next, to determine $p_{kj}$, the condition \eqref{eqn:intra-if-1R} and also a relabelled version of \eqref{eqn:intra-if-1R}: \mbox{$x^{p_{ij}} \nequiv x^{p_{kj}}$} must be satisfied.
  Since $p_{ji}$ and $p_{ki}$ are already fixed, there is only one solution left for $p_{kj}$.
  Then, the remaining parameter $p_{jk}$ can be determined uniquely by~\eqref{eqn:IA-1R} for relabelled indices, i.e., \mbox{$e_{\mathrm{R}k} x^{p_{jk}} \equiv e_{\mathrm{R}j} x^{p_{kj}}$}.
  The condition \eqref{eqn:mac-if-1R} holds, since these interferences are pair-wise aligned:
  \begin{align*}
               e_{\Rx i}x^{p_{ji}} &\nequiv e_{\Rx k} x^{p_{jk}} \,\mod(x^3-1)\\
   \Rightarrow e_{\Rx j}x^{p_{ij}} &\nequiv e_{\Rx j} x^{p_{kj}} \,\mod(x^3-1)\\
   \Rightarrow          x^{p_{ij}} &\nequiv x^{p_{kj}} \,\mod(x^3-1),
  \end{align*}
  yielding a relabelled version of \eqref{eqn:intra-if-1R} that already holds.
  Using similar steps, \eqref{eqn:inter-if-1R} holds as well.
  The received signal at $\Rx$ can be expressed as:
  \begin{align*}
   r_{\Rx}(x) \equiv &\ \ \ (W_{ij}+W_{ji})e_{\Rx i} x^{p_{ji}}+(W_{jk}+W_{kj})e_{\Rx j} x^{p_{kj}} \\
    &+(W_{ik}+W_{ki})e_{\Rx k} x^{p_{ik}} \,\mod(x^3-1).
  \end{align*}
  In the second hop, each transceiver $\Tx_j$ applies SIC to remove the self-interference $W_{ij}$ and $W_{kj}$ \resp, so that the dedicated messages $W_{ij}$ and $W_{kj}$ can be decoded interference-free.
  The interference contained in the remaining dimension is discarded.
  Cyclic IA always achieves the upper bound for arbitrary~$\boldsymbol{e}$ and~$\boldsymbol{f}$.~$\blacksquare$
\end{itemize}

\subsection{General Case}
For the $Y$-\,channel with the general messaging matrix in \eqref{eqn:msg-matrix}, the signals are aligned in pairs~by:
\begin{align}
\label{eqn:IA-1R-gen}
 e_{\mathrm{R}i} x^{p_{ji}^{[m]}} \equiv e_{\mathrm{R}j} x^{p_{ij}^{[m]}}\,\modxn.
\end{align}
with \mbox{$m \in \{1,\dots,\min(\alpha_{ji},\alpha_{ij})\}$}.
The remaining signals get an own dimension and thus are transmitted in an ordinary multiple-access scheme.

\begin{theorem}
On the $Y$-\,channel, Cyclic~IA with SIC achieves:
 \begin{align*}
  \DoF \leq \frac{\alpha_{12}+\alpha_{21}+\alpha_{13}+\alpha_{31}+\alpha_{23}+\alpha_{32}}{\max(n_1,n_2,n_3)} \leq 2,
 \end{align*}
 within $n \geq \max(n_1,n_2,n_3)$ dimensions for distinct \mbox{$i,j,k \in \mathcal{K}$}:
 \begin{align*}
  n_j &= \alpha_{ji}+\alpha_{jk}+\max(\alpha_{ik},\alpha_{ki}).
 \end{align*}
\end{theorem}
\emph{Proof:}
\begin{itemize}
 \item [(a)] \emph{Necessity of $n \geq \max(n_1,n_2,n_3)$ dimensions:} \newline
  At user $\Tx_j$ a number of $\alpha_{ji} + \alpha_{jk}$ dimensions is necessary to satisfy the multiple-access interference condition.
  To satisfy the inter-user interference conditions at $\Tx_j$, inter-user interference must be aligned within $\max(\alpha_{ik},\alpha_{ki})$ dimensions.
  Relay $\Rx$, does not impose further constraints and demands $n \geq \max(n_1,n_2,n_3)$ dimensions.
 \item [(b)] \emph{Sufficiency of Cyclic IA and SIC:} \newline
  The Cyclic~IA scheme given in Lem.\,\ref{lem:Y-perfIA} can also be generalized to an arbitrary number of submessages.
  This proof is essentially analogous to Thm.~\ref{thm:IA-FD-arb}. 
  Each pair of aligned signals occupies exactly one dimension at each receiver, either as a dedicated signal, or as interference.
  Thus $\min(\alpha_{ij},\alpha_{ij})$ dimensions are used for the aligned signals from $\Tx_i$ and $\Tx_j$ at $\Tx_k$.
  The remaining $|\alpha_{ij}-\alpha_{ji}|$ submessages per message are separately allocated to $|\alpha_{ij}-\alpha_{ji}|$ dimensions by multiple-access, so that $n_j$ dimensions are needed at $\Tx_j$.
  All submessages are conveyed interference-free within $n$ dimensions.
  There are no requirements on channel vector~$\boldsymbol{e}$.~$\blacksquare$
\end{itemize}

\subsection{$Y$\,-\,$\Delta$ Product}
\label{sec:equivalence}
Interestingly, the achieved DoF of the considered $3$-way $\Delta$ channel and the \mbox{$Y$-\,channel} coincide exactly, even for an arbitrary number of submessages.
This observation motivates the idea that these two channels are convertible to each other from a DoF perspective, similar to a $Y$\,-\,$\Delta$ transformation of an elementary electrical circuit with resistors.
The outer product of the uplink and downlink channel vectors $\boldsymbol{e}$ and~$\boldsymbol{f}$ of the $Y$-\,channel provides an effective channel matrix~$\boldsymbol{D}$ of the \mbox{$3$-way $\Delta$} channel:
\begin{align}
 \label{eqn:decomposition}
 \boldsymbol{D} = \boldsymbol{e}^\transpose \boldsymbol{f} = \boldsymbol{f}^\transpose \boldsymbol{e}.
\end{align}
Although, the $Y$\,-\,$\Delta$ product representation is valid for any \mbox{$Y$-\,channel}, it is not always valid for any $3$-way $\Delta$ channel, since a matrix decomposition of \eqref{eqn:decomposition} is not always available for arbitrary~$\boldsymbol{D}$.
The $Y$-\,channel is considered as a special case of the \mbox{$3$-way $\Delta$} channel as the separability conditions of these two channels are equivalent to their respective counter-part and interestingly, SA translates to IA in the product representation.
A particular gain of this representation is that the upper bound of \eqref{eqn:obound-TW} also carries over to the \mbox{$Y$-\,channel} when including SIC.

It is yet an open question whether a generalization for \mbox{$K$-\,user} multi-way (or $K$-way) networks and $K$-user \mbox{$Y$-\,networks} would lead to a corresponding star-mesh product.

\section{Relation to Signal-Space IA}
\label{sec:mimo-dof}

\subsection{Three-Way Channel with Time-Varying Coefficients}
In \cite[Thm.~2]{318}, a fully-connected full-duplex multi-way channel with $K$ users and time-varying channel coefficients is considered.
The particular case of $K=3$ users yields an upper bound of $\frac{K(K-1)}{2K-3}=2$ DoF.

To the best of our knowledge, upper bounds and corresponding achievable schemes for a MIMO $3$-way $\Delta$ channel with an arbitrary number of transmit/receive antennas at the transceivers are not provided by current literature yet.
In this light, our proposed schemes are intended to provide some new conceptual insights to these yet unsolved problems.

\subsection{MIMO $Y$-\,Channel}
For signal-space IA in MIMO systems \cite{539,582}, the DoF are basically characterized by the bottleneck in terms of the number of antennas per transceiver $\Tx_j$ and per relay $\Rx$.
For the $Y$-\,channel, the users $\Tx_j$ also use a number of $A_{\Tx_j}$ full-duplex antennas and the relay $\Rx$ uses a number of $A_{\Rx}$ full-duplex antennas.
By utilizing Interference Nulling Beamforming \cite{539,582} at~$\Rx$, those particular dimensions at the transceivers, that are reserved for receiving aligned interference, may be omitted if the interference forwarded from~$\Rx$ can be beamformed into the nullspace of the transceivers.
The transceivers omit these reserved dimensions by reducing their number of antennas.

By assuming \Wlog that $A_{\Tx_1} \geq A_{\Tx_2} \geq A_{\Tx_3}$, it is shown by Chaaban \etal in \cite{582}, that the upper bound on the number~of:
\begin{align*}
\DoF_{\mathrm{MIMO}} \leq \min(2 A_{\Tx_2}+2 A_{\Tx_3}, A_{\Tx_1} + A_{\Tx_2} + A_{\Tx_3}, 2 A_{\Rx}) 
\end{align*}
is achievable.
In all cases, perfect IA is applied to maximize the achievable DoF.
To relate $DoF_\mathrm{MIMO}$ to the given $\DoF$ measure, we normalize it by the minimal necessary~$A_{\Rx}$:
\begin{align*}
 \DoF = \frac{\DoF_{\mathrm{MIMO}}}{A_{\mathrm{R}}}.
\end{align*}
Then, each case clearly corresponds to the maximal number of $2$~DoF for perfect~IA as shown in the previous section:
\newline
\emph{Case A). $\DoF_{\mathrm{MIMO}} \leq 2 A_{\Tx_2}+2 A_{\Tx_3}$:} \newline
  As given in the first case of \cite[Sect.~IV-A]{582}, $A_{\Rx}=A_{\Tx_2}+A_{\Tx_3}$ antennas are used at $\Rx$.
  The normalized DoF yield:
  \begin{align*}
   \DoF \leq \frac{2A_{\Tx_2}+2A_{\Tx_3}}{A_{\Rx}}=2.
  \end{align*}
\emph{Case B). $\DoF_{\mathrm{MIMO}} \leq A_{\Tx_1} + A_{\Tx_2} + A_{\Tx_3}$:} \newline
  In the second case, $A_{\Rx} = \frac{1}{2}(A_{\Tx_1} + A_{\Tx_2} + A_{\Tx_3})$ antennas are necessary, as given in \cite[Sect.~IV-B]{582}.
  The normalized DoF~are:
\begin{align*}
 \DoF \leq \frac{A_{\Tx_1}+A_{\Tx_2}+A_{\Tx_3}}{A_{\Rx}} = 2.
\end{align*}
\emph{Case C). $\DoF_{\mathrm{MIMO}} \leq 2 A_{\Rx}$:} \newline
In the last case, \mbox{$A_{\Rx} \leq \min(A_{\Tx_2}+A_{\Tx_3},\frac{1}{2}(A_{\Tx_1}+A_{\Tx_2}+A_{\Tx_3}))$} is assumed in \cite[Sect.~IV-C]{582}, and the normalized DoF yield:
\begin{align*}
 \DoF \leq \frac{2 A_{\Rx}}{A_{\Rx}}=2.
\end{align*}
All three cases achieve the same normalized DoF of $2$ and also include the result by Lee \etal in \cite{539}.

\bibliographystyle{IEEEtran}
\bibliography{paper}

\end{document}